\documentstyle[12pt]{article}

\newcommand{\EQ}{\begin{equation}}
\newcommand{\EN}{\end{equation}}
\newcommand{\bea}{\begin{eqnarray}}
\newcommand{\ena}{\end{eqnarray}}
\newcommand{\bdis}{\begin{displaymath}}
\newcommand{\edis}{\end{displaymath}}
\newcommand{\vs}[1]{\vspace{#1 mm}}

\renewcommand{\a}{\alpha}
\renewcommand{\b}{\beta}
\renewcommand{\c}{\gamma}
\renewcommand{\d}{\delta}
\renewcommand{\v}{\Delta}

\renewcommand{\o}{\omega}
\renewcommand{\t}{\tau}

\newcommand{\HFP}{H_{FP}}
\newcommand{\tHFP}{{\tilde H}_{FP}}

\newcommand{\hA}{\hat A}
\newcommand{\hp}{\hat \pi}

\newcommand{\pa}{\partial}
\newcommand{\limt}{\lim_{\t \rightarrow \infty}}
\newcommand{\itd}{\int \! d^3x}

\newcommand{\igr}{\int \! {\cal D}A}

\newcommand{\dpi}{{\d \over \d A^{in}}}
\newcommand{\hi}{{\hat i}}
\newcommand{\hj}{{\hat j}}
\newcommand{\U}{\sqcap_{i\neq j} U}
\newcommand{\nn}{\nonumber \\}

\begin{document}

\topmargin 0pt
\oddsidemargin 5mm

\newcommand{\NP}[1]{Nucl.\ Phys.\ {\bf #1}}
\newcommand{\PL}[1]{Phys.\ Lett.\ {\bf #1}}
\newcommand{\CMP}[1]{Comm.\ Math.\ Phys.\ {\bf #1}}
\newcommand{\PR}[1]{Phys.\ Rev.\ {\bf #1}}
\newcommand{\PRL}[1]{Phys.\ Rev.\ Lett.\ {\bf #1}}
\newcommand{\PREP}[1]{Phys.\ Rep.\ {\bf #1}}
\newcommand{\PTP}[1]{Prog.\ Theor.\ Phys.\ {\bf #1}}
\newcommand{\PTPS}[1]{Prog.\ Theor.\ Phys.\ Suppl.\ {\bf #1}}
\newcommand{\NC}[1]{Nuovo.\ Cim.\ {\bf #1}}
\newcommand{\JPSJ}[1]{J.\ Phys.\ Soc.\ Japan\ {\bf #1}}
\newcommand{\MPL}[1]{Mod.\ Phys.\ Lett.\ {\bf #1}}
\newcommand{\IJMP}[1]{Int.\ Jour.\ Mod.\ Phys.\ {\bf #1}}
\newcommand{\AP}[1]{Ann.\ Phys.\ {\bf #1}}
\newcommand{\RMP}[1]{Rev.\ Mod.\ Phys.\ {\bf #1}}
\newcommand{\PMI}[1]{Publ.\ Math.\ IHES\ {\bf #1}}
\newcommand{\JETP}[1]{Sov.\ Phys.\ J.E.T.P.\ {\bf #1}}
\newcommand{\TOP}[1]{Topology\ {\bf #1}}
\newcommand{\AM}[1]{Ann.\ Math.\ {\bf #1}}
\newcommand{\ZP}[1]{Z. \ Phys.\ {\bf #1}}
\newcommand{\LMP}[1]{Lett.\ Math.\ Phys.\ {\bf #1}}
\newcommand{\CRASP}[1]{C.R.\ Acad.\ Sci.\ Paris\ {\bf #1}}
\newcommand{\JDG}[1]{J.\ Diff.\ Geom.\ {\bf #1}}
\newcommand{\JSP}[1]{J.\ Stat.\ Phys.\ {\bf #1}}

\begin{titlepage}
\setcounter{page}{0}
\begin{flushright}
NBI-HE-95-27\\
August 1995\\
gr-qc/9508045\\
\end{flushright}

\vs{8}
\begin{center}
{\Large Lattice Quantum Gravity from Stochastic 3-Geometries}

\vs{15}
{\large Naohito Nakazawa\footnote{On leave of absence from
Department of Physics, Faculty of Science, Shimane University,
Matsue 690, Japan. \ e-mail address:
nakazawa@nbivax.nbi.dk, nakazawa@ps1.yukawa.kyoto-u.ac.jp}}\\
{\em The Niels Bohr Institute, Blegdamsvej 17, DK-2100 Copenhagen \O,
Denmark \\}

\end{center}

\vs{8}
\centerline{{\bf{Abstract}}}

I propose the Langevin equation for 3-geometries in the Ashtekar's
formalism to describe 4D Euclidean quantum gravity, in
the sense that the corresponding Fokker-Planck hamiltonian recovers
the hamiltonian in 4D quantum gravity exactly.
The stochastic time corresponds to the
Euclidean time in the gauge,
$N=1$ and $N^i=0$. In this approach, the time evolution in
4D quantum gravity is understood
as a stochastic process where the quantum fluctuation
of \lq \lq triad \rq\rq
is characterized by the curvature at the one unit time step before.
The lattice regularization of 4D quantum gravity is presented in this
context.

\end{titlepage}
\newpage
\renewcommand{\thefootnote}{\arabic{footnote}}
\setcounter{footnote}{0}

In the course to study $QCD_4$ in terms of the
Nicoli-Langevin maps~\cite{PS}\cite{Ni},
it is shown that stochastic quantization~\cite{PW} of
3D Chern-Simons theory recovers the time
evolution in 4D Euclidean Yang-Mills theory where the stochastic
( or fictitious ) time is
interpreted as the Euclidean time~\cite{CH}.
There is also another example, found in stochastic quantization of
matrix models with loop space hamiltonian
at the double scaling limit, that the stochastic
time corresponds to the Euclidean time coordinate in
2D quantum gravity~\cite{JR}\cite{Na1}.
These facts motivate us to interpret the stochastic time as the
Euclidean time coordinate even in 4D quantum gravity.
In this short note, I point out that the time evolution
in 4D quantum gravity
is described by a Langevin equation for 3-geometries in terms of
the Ashtekar's canonical field variables~\cite{Ash}\cite{JS} by showing that
the corresponding
Fokker-Planck hamiltonian operator recovers the hamiltonian of 4D Euclidean
quantum gravity exactly~\cite{Mos}.
The stochastic time
is interpreted as the Euclidean time with
the gauge $N=1$ and $N^i = 0$, where $N$ and $N^i$ are the lapse function
and the shift vector, respectively.
The Hartle-Hawking type boundary condition
is naturally imposed in this scheme by specifying the initial probability
distribution functional.
Then I propose the lattice regularization of this approach with 3D cubic
lattice. The corresponding F-P hamiltonian defines the lattice
regularization for the hamiltonian of 4D Euclidean quantum gravity in
Ashtekar's formalism.

At first, I introduce the basic Langevin equation for 3-geometries in
the Ashtekar's variables to recover the
hamiltonian of 4D quantum gravity with the corresponding
F-P hamiltonian defined latter.
The simplest form of the Langevin equation
is defined by,
\bea
\v A^a_i(x, \t)
&=&  \v\zeta^a_i(x, \t)                          \ , \nn
<\v\zeta^a_i(x, \t) \v\zeta^b_j(y, \t)>_\zeta
&=& {\kappa\over 2} \v\t \epsilon^{abc}<F^c_{ij}(x, \t)>_\zeta\d^3(x-y)
              \ ,
\ena
where $A^a_i (x, \t)$ is a SU(2) gauge field in Euclidean
Ashtekar's canonical
formalism\footnote{
I use the notation,
$
F^a_{ij} = \pa_i A^a_j - \pa_j A^a_i + \epsilon^{abc}A^b_i A^c_j  \ ,
$
and
$
D^{ac}_i = \d^{ac}\pa_i + \epsilon^{abc}A^b_i   \ .
$
The field variable $A^a_i (x)$ is $real$ in the Euclidean
Ashtekar's formalism.
}. In this note, Latin indices \lq\lq i,j,k,...\rq\rq denote
the spatial part of the spacetime coordinate indices.
While the Latin letters
\lq\lq a,b,c,...\rq\rq denote the spatial part of the internal indices.
$x, y,...$ denote spatial spacetime coordinates.
The one step time evolution is defined by
$\v A^a_i(x, \t) \equiv A^a_i(x, \t + \v\t ) - A^a_i(x, \t)$ in (1).
$\v\t$ is
the unit of the discretized stochastic time and
$\t$ denotes the stochastic time after $n$ steps, $\t \equiv
n\v\t$.
The discretization of the stochastic time is considered
for convenience of stochastic calculus and
for understanding the precise meaning of the noise correlation.
The coupling constant $\kappa$ is defined by
$\kappa \equiv 16\pi G$ with $G$, the gravitational ( Newton's ) constant
in the natural unit $\hbar = c = 1$.
The noise variable in (1) is not a simple white noise.
The expectation value of the R.H.S. of
the noise correlation is understood to be taken with respect to the noises
up to the one unit time step before, $\t - \v\t$, in the sense of
Ito's stochastic calculus~\cite{Ito}. More precisely,
if the expectation value is taken only at a specified
stochastic time, $\t$, then
the R.H.S. of the noise correlation is not the
expectation value since it does not depend on the noise at $\t$.
It is also equivalent to require $< \v\zeta^a_i (\t) > = 0$ in Ito's
calculus.

The basic Langevin equation is manifestly covariant under the
SU(2) local gauge transformation,
$
A^a_i (x) \rightarrow A^a_i + D^{ab}_i \o^b (x)       \ .
$
While it is not covariant under the spatial general coordinate
transformation,
$
A^a_i (x) \rightarrow A^a_i + F^a_{ji} \xi^j (x)       \ ,
$
due to the divergent term,
$
\kappa \v\t \d^3(0) F^a_{ji}\xi^j        \ ,
$
which appears in the transformation of $\v A^a_i$.
This divergent term is formally ( but not well-defined ) cancelled by
adding a term in the R.H.S. of the Langevin equation (1).  The extra
term comes
from the invariant path-integral measure as we will see later.

In terms of the solution of the Langevin equation, the following
equality holds.
\bea
<\Pi_{x,i,a}\d \big( A^a_{i\ \zeta} (x, \t)
&-& A^{a(final)}_i (x)\big) >_\zeta                      \nn
&=& <A^{(final)}|{\rm e}^{-\t \tHFP [\hp, \hA ]}|A^{(initial)}>     \ , \nn
&=& <A^{(initial)}|{\rm e}^{-\t \HFP [\hA, \hp ]}|A^{(final)}>         \ .
\ena
In the L.H.S., $A^a_{i \zeta} (x, \t)$ denotes the solution of
the Langevin equation with the initial condition,
$
A^a_i (x, 0) = A^{a(initial)}_i (x)         \ .
$
In the R.H.S., $\tHFP [\hp, \hA]$ is defined by,
\bea
\tHFP [ \hp, \hA ]
&=& - {\kappa\over 4} \itd \epsilon^{abc} \hp_a^i (x) \hp_b^j (x)
 {\hat F}^c_{ij}(x)             \ ,  \nn
\HFP [\hA, \hp ]
&=& - {\kappa\over 4} \itd
\epsilon^{abc} {\hat F}^c_{ij}(x)
    \hp_a^i (x) \hp_b^j (x)            \ .
\ena
To show the equality (2), the commutation relation,
\EQ
[ \hp_a^i (x) \ , \hA^b_j (y) ]
= \d_a^b \d^i_j\d^3 (x-y)              \ ,
\EN
and the vacuum, $|0>$ with
$\hp^i_a |0> = <0|\hA^b_j = 0$, are assumed. The representation
of the states is given by,
\bea
<A|
&\equiv& <0|\ {\rm exp}[\itd A^a_i(x) \hp_a^i (x)]          \ , \nn
|A>
&=& \Pi_{x,a,i}\d \big( \hA^a_i(x) - A^a_i (x) \big) |0>      \ .
\ena
The Fokker-Planck hamiltonians (3) are just the hamiltonian
for 4D quantum gravity without cosmological
term in Ashtekar's variables~\cite{Ash} with different operator
orderings\footnote{
For a detailed derivation of the F-P hamiltonian operator from the Langevin
equation, see also Ref.~\cite{Na1}.
}.
The annihilation operator $\hp_a^i$
is interpreted as the
\lq\lq triad \rq\rq in the Euclidean Ashtekar's formalism.
Namely\footnote{
$
{\tilde e}_a^i \equiv q^{1/2}e_a^i   \ ,
$
where
$
q = {\rm det}(q_{ij})   \
$
with spatial metric
$
q^{ij} = e_a^i e_a^j   \ .
$
},
$
\hp^j_a (x) = - {2 i\over \kappa}{\tilde e}^j_a (x)  \ .
$
The stochastic time evolution with the Langevin equation (1)
corresponds to the Euclidean time evolution in 4D quantum gravity
with the gauge fixing, $N=1$ and $N^i = 0$\footnote{
In 2D quantum gravity, it is observed that this gauge fixing
recovers the time evolution defined in non-critical string field
theories\cite{FIKN}.
}. In the stochastic process (1),
the \lq\lq triad \rq\rq plays the role of the noise variable and the
\lq\lq quantum fluctuation \rq\rq of the triad is characterized
by the curvature.
Though the two expressions in (2) are precisely equivalent,
the second
definition of the F-P hamiltonian in (3) has an advantage
for further discussion.
This is because the vacuum satisfies the ( local ) hamiltonian
constraint with the operator ordering in $\HFP$,
$
<0| {\cal H}( \hA (x), \hp (x) ) = {\cal H}(\hA (x), \hp (x)) |0> = 0   \ ,
$
where
$
H_{FP}[\hA, \hp] \equiv \itd {\cal H}(\hA (x), \hp (x))  \ .
$
It should be noticed that the
operator
ordering of the
F-P hamiltonian operator $\HFP$ is deferent from that appeared in the F-P
equation for the probability distribution.

Let us consider the initial distribution dependence of the probability
distribution functional by averaging the expectation value
(2) with respect to the initial probability distribution. It is defined
by integrating out the initial configuration $A^{a(initial)}_i (x)$,
on which the solution
of the Langevin equation $A^a_{i \zeta} (x, \t)$ depends,
with the distribution, $P[A^{(initial)}, 0]$, in the L.H.S. of (2).
The integration of $A^{a(initial)}_i$ gives a generalized
form of the distribution
functional $P[A, \t]$, which is defined by
$<O[A_\zeta] (\t)>_\zeta \equiv \igr O[A]P[A, \t]$, as follows
\bea
P[A, \t]
= \int\!{\cal D}A^{initial}
<A^{initial}|P[\hA, 0]{\rm e}^{-\t \HFP [\hA, \hp]}|A>     \ .
\ena
For an arbitrary observable $O[A]$, the average with respect to the
initial value distribution also gives,
\bea
<O[A_\zeta (\t)]>_\zeta
= \int\!{\cal D}A^{initial}
<A^{initial}|P[\hA, 0]{\rm e}^{-\t \HFP [\hA. \hp]}O[\hA ]|0>     \ .
\ena
In the definition of the expectation value in the L.H.S. of (7),
the average is
also taken with respect to the initial values with the distribution
$P[A^{initial}, 0]$. For example,
eq.(2) is given for
$O[A] = \Pi_{x, a, i} \d \Big( A^a_i(x) - A^{a(final)}_i (x) \Big)$, with
the initial distribution,
$P[A, 0] = \Pi_{x, a, i} \d \Big( A^a_i(x) - A^{a(initial)}_i (x) \Big)$.

Eq.(7) leads the time evolution equation for the expectation values of
observables,
\bea
{d \over d\t}<O[A_\zeta (x, \t)]>_\zeta
= - <H_{FP}[ A^b_j (x), { \d \over \d A^b_j(x)} ]
O[A^a_i(x)] {\big|}_{A^a_i (x) = A^a_{i{}\zeta} (x, \t)}>_\zeta   \ ,
\ena
as it is expected from the Langevin equation (1).
I here notice that
the initial condition dependence in (7) implies a constraint
for initial distribution itself due to the equation,
\bea
{d \over d\t}<O[A_\zeta (x, \t)]>_\zeta
= - \int\!{\cal D}A^{in} P[A^{in}, 0]
\HFP[A^{in}, \dpi ]<A^{in}|{\rm e}^{-\t \HFP [\hA, \hp]}O[\hA]|A>     \ .
\ena
The existence of the equilibrium limit requires that
the R.H.S. in (9) should be zero at the infinite stochastic time.
A trivial solution of
this
constraint is the vanishing curvature at any spatial points,
$
P[A, 0] = \Pi_{x, a, ij} \d \Big( F^a_{ij}(x) \Big)        \ .
$
This initial distribution, however, should be excluded
because the R.H.S. of (8) is identically
zero even at finite
stochastic time and there is no time development.
In general, if we choose a solution of
the hamiltonian constraint as
the initial value distribution, obviously there is no time evolution in the
Fokker-Planck type equation (8).
Especially, the constraint does not allow us to solve the Langevin
equation with the initial condition $A^{a(initial)}_i (x) = 0$ for the
existence of the time evolution.

To specify
the physical boundary condition and
to study a class of
solutions for the hamiltonian constraint of 4D quantum gravity in this
context, we may choose the initial condition which generates a
nontrivial time evolution
\bea
P_{H-H}[A, 0] = \Pi_{x \neq z_0, a, i} \d \Big( F^a_{ij}(x) )
\Big)        \  .
\ena
The spacetime point, $x=z_0$ at $\t = 0$, is identified to the point where the
3D spatial manifold is absorbed into nothing
in an analogous sense of the Hartle-Hawking type boundary condition~\cite{HH}.

As it is clear from (7), initial distributions and
observables are the key quantities to specify 3-boundaries on quantized
4D spacetime manifolds in this approach.
These quantities are the solution of the momentum constraint
and the SU(2) Gauss law constraint.
There are some candidates which may be useful to
characterize 3-boundaries, such as the extrinsic curvature term,
3D Chern-Simons term,
topological invariants~\cite{Wi} and loop variables~\cite{Smo} for
which the present formalism can be applied. Under the appropriate
choice of these observables, one can show
the gauge invariance ( gauge independence ) of the expectation value of
observables, following the standard method to
fix the gauge in the Langevin equation~\cite{Zw}\cite{Na2}.
For spatial general coordinate invariance and SU(2) gauge invariance,
the gauge fixed Langevin equation is
given by,
\bea
\v A^a_i(x, \t)
=  - \v\t \big\{ D^{ab}_i \Psi^b + F^a_{ij}\Phi^j  \big\} (x, \t)+
\v\zeta^a_i(x, \t)             \ ,
\ena
with the same noise correlation as eq.(1). The gauge fixing
corresponds to choose the multiplier fields,
$\Psi^a (x)$ and $\Phi^i (x)$, as specified functionals of $A^a_i$'s.
Then, the F-P hamiltonians are given by,
\bea
\tHFP^{(g.f.)}[\hp, \hA ]
&=& \tHFP + \itd \Big\{ \Phi^i(x) {\tilde C}^M_i (x)
+ \Psi^a (x){\tilde C}^a_G (x)
\Big\}         \ , \nn
\HFP^{(g.f.)}[\hA, \hp ]
&=& \HFP - \itd \Big\{ \Phi^i(x) C^M_i (x) + \Psi^a (x) C^a_G (x)
\Big\}                             \ ,
\ena
instead of eq.(3), with
\bea
C^M_i (x) &\equiv&  {\hat F}^a_{ij} \hp_a^j              \ , \nn
C^a_G (x) &\equiv&  {\hat D}^{ab}_i \hp_b^i             \ .
\ena
$C^M_j$ and $C^a_G$ defines the momentum constraint and the Gauss law
constraint, respectively.
In the first expression in (12),
${\tilde C^M_j}$ and ${\tilde C_G^a}$, are defined
by replacing the operator orderings in the original constraints so that
the conjugate momentum, $\hp_b^j$'s, are gathered to the left to
$A^a_i$'s in the constraints.
The probability distribution and the expectation value defined
by the Langevin equation (11) are given
by eqs.(6) (7) together with the F-P hamiltonian (12) instead of (3).
As it is clear from (7) with the second expression of the
F-P hamiltonian in (12), the insertion of constraints
does not change the expectation value of the observables
provided that the observables are the solution of these
constraints, such as the extrinsic curvature term and 3D Chern-Simons
term.
We notice that the operator ordering in the second expression in
(12) is essential for the local gauge
invariance in our definition of the vacuum, namely
the vacuum is gauge invariant,
$
<0|C^M_i = C^M_i|0> = <0|C^a_G = C^a_G|0> = 0  \ ,
$
with the operator ordering in (13).
Therefore, (11) gives
the same expectation value for arbitrary observables as one given by
(1), provided that these observables are
the solution of the constraints in (13). The gauge fixing
in (1) corresponds to $N=1$, $\Phi^i = \Psi^a = 0$, where $\Phi^i$ is
essentially the shift vector.
On the other hand, as for the partition function defined by the Langevin
equation (11), which is given by (6)
with (12), the gauge fixing procedure leads a gauge invariant
( more precisely BRS invariant ) partition function with a suitable
choice of multiplier field, $\Phi^i$ and $\Psi^a$, as functionals of
dynamical variables. To fix the gauge more generally for the
reparametrization invariance of the time coordinate, one has to
introduce the lapse function by multiplying the R.H.S. of the noise
correlation (1) with it and choose it as a specified functional of
$A^a_i$'s. Though the problem is not addressed in this note explicitly,
it is possible following the general framework~\cite{Na2}.

Now let us consider the regularization procedure in this context.
As it has been noticed in (1), an extra term is
necessary for the covariance of the Langevin equation under the
spatial general coordinate transformation.  It is a direct consequence
of the fact that the Ito's calculus picks up the Jacobian factor which
comes from the change of variables in the path-integral
measure~\cite{Gr}\cite{Na3}.
One way
to specify the path-integral measure in configuration space
is to introduce a regularization for the noise
correlation in (1).
I here propose a lattice
regularization of the Langevin equation and the noise correlation (1)
in which the invariant
property of the regularized Langevin equation in the sense of Ito's calculus
naturally introduces the invariant path-integral measure in the configuration
space of link variables.

The lattice regularized Langevin equation is defined by
\bea
\v U_l (\t )
&=& \Big\{ \v W_l (\t ) +
\v\t |{\cal G}|^{1/2} (D_L)^b_{l'}
\Big( |{\cal G}|^{-1/2} {\cal G}^{ab}_{ll'} \Big)T^a \Big\} U_l (\t)   \ \nn
| {\cal G}|
&\equiv& {\rm det}( {\cal G}^{ab}_{ll'} )         \ . \nn
\ena

Here the dynamical variable and the noise variable,
$U_l$ and $\v W_l$ respectively, have been assigned
on the link, $l \equiv (x, i)$, of 3-dimensional cubic lattice,
which is specified by the site $x$
and its nearest neighbor in the i-th direction $x + \hi$.
$U_l$ is an element of SU(2) group in the adjoint
representation. While the noise, $\v W_l$ is algebra-valued in
the adjoint representation,
$\v W_l = \v W^a_l T^a$, with $[T^a \ , T^b] =\epsilon^{abc}T^c$.
$(D_L)^a_l$ is the left Lie derivative
on the group manifold defined by
\bea
[ ( D_L )^a_l, \  U_{l'}]
 = \d^{ll'} T^a U_l                   \ , \quad
[ ( D_L )^a_l, \ (D_L)^b_l ]
 = - \epsilon^{abc}(D_L)^c_l            \ .
\ena
The regularized Langevin equation describes the one step time
evolution of the link variables,
$
U_l(\t + \v\t ) = U_l(\t ) + \v U_l (\t )       \ .
$
It should be noticed that $U_l (\t + \v\t)$ is also an element of
SU(2) group.
The quantity
$
{\cal G}^{ab}_{ll'} \ ,
$
is interpreted as the inverse of the \lq\lq superspace \rq\rq metric,
here the superspace is
spanned by the link variables $\{ U_l \}$.
The inverse of the superspace metric is given in the following
regularized correlation of the noises defined on the links,
$l=(x,i)$ and $l'=(y,j)$.
\bea
<(\v W_l)_{\a\b}(\t) (\v W_{l'})_{\c\d} (\t)>
= \v\t <{\cal G}^{ab}_{ll'}(\t)T^a_{\a\b}T^b_{\c\d}>      \ ,
\ena
where,
\bea
{\cal G}^{ab}_{ll'}(\t)T^a_{\a\b}T^b_{\c\d}
&=& g_0\d_{x, y} \Big\{ \d_{\a\d} (\U)_{\b\c} + \d_{\b\c}(\U)_{\a\d}
- \d_{\b\d} (\U)_{\a\c} - \d_{\a\c}(\U)_{\b\d} \Big\}            \ , \nn
(\U)_{\a\c}
&\equiv& {1\over 2}\Big\{
\big( U(x, i)U(x+\hi, j)U(x+\hj, i)^\dagger U(x, j)^\dagger \big)_{\a\c} \nn
&-& \big(  U(x, j)U(x+\hj, i)U(x+\hi, j)^\dagger U(x, i)^\dagger
 \big)_{\a\c} \Big\}                  \ .
\ena
$g_0$ is an arbitrary constant.

The lattice regularization has been determined from the following
requirements. (i) The Langevin equation (14) and the noise correlation (16)
should be covariant under the SU(2) local
gauge transformation,
\bea
U(x, i)
&\rightarrow& V^{-1}(x)U(x,i)V(x+\hi)       \ , \nn
\v W(x, i)
&\rightarrow& V^{-1}(x)\v W(x,i)V(x)       \ .
\ena
(ii) The model recovers the Langevin equation and the noise correlation
in (1) at the naive continuum limit defied by,
$U(x, i) \approx 1 + A^b_i T^b a$ and $\v W(x, i) \approx \v\zeta^b_i T^b a$.
At the limit,
the coupling constant and the stochastic time are scaled,
$g_0 a^2 \rightarrow {\kappa\over 2}$
and $\v \t \rightarrow  a^{-1} \v \t $, as
the lattice spacing $a$ goes to zero, $a \rightarrow 0$.
(iii)
The regularized Langevin equation (14) is covariant under the
\lq\lq general coordinate transformation \rq\rq in superspace,
$\{ U(x, i) \}$.
\bea
U(x, i)_{\a\b} \rightarrow V(x, i)_{\a\b}[U]     \ ,
\ena
where $V(x,i)$ is also an element of SU(2) group in the adjoint
representation and an
arbitrary functional of $U(y,j)$. The second term in
the R.H.S. in (14) is
necessary for the invariance in Ito's calculus~\cite{Gr}. The
requirement (iii) is introduced to realize the spatial coordinate invariance
in the lattice regularization.
By using these requirements, it is straightforward to construct
the model in the fundamental representation.  It is given by simply
replacing the
inverse of the superspace metric (17) to the following one.
\bea
{\cal G}^{ab}_{ll'}(\t)T^a_{\a\b}T^b_{\c\d}
= g'_0\d_{x, y} \Big\{  \d_{\b\c}(\U)_{\a\d} - \d_{\a\d} (\U)_{\c\b}
\Big\}            \ .
\ena

By applying the argument in Ref.~\cite{Gr}, I have introduced the second term
in the R.H.S. of (14) for the covariance of the lattice regularized
Langevin equation under the transformation (19).
The similar argument is formally possible even
in the non-regularized
version (1) if we identify the R.H.S. of the
noise correlation in (1) to the inverse of
the continuum superspace metric.  Indeed, from the requirement (ii),
the naive continuum limit of these equations, (14) (15) and (16),
coincides with those in
eq.(1) except a term which comes from
the second term in the R.H.S. of the Langevin equation (14).
The extra term has the same structure as the second term
in the regularized Langevin
equation (14) if we
identify the noise
correlation in (1) as the inverse of the superspace
metric in Ashtekar's configuration variables.
The extra contribution which comes
from this term
under the spatial general coordinate transformation
formally cancels the divergent term, $\kappa\v\t \d^3(0)F^a_{ji}\xi^j$,
which appears in the transformation in
$\v A^a_i$. The extra term appeared in the naive
continuum limit of (14) actually
represents the contribution from the path-integral measure
and it is necessary for the invariant property of the Langevin
equation (1) in a formal
sense, however, it is not well-defined without regularization.
A more detailed discussion, especially
on the covariant property of the lattice regularization
under the spatial general coordinate
transformation in the present approach, will be published in
elsewhere~\cite{Napre}.

By using the same method developed in the continuum version,
one can derive the regularized version of the expectation value,
such as (6). I only comment on the three important
consequences of the lattice regularized
Langevin equation (14). The first is
the corresponding lattice regularized F-P hamiltonian which is given
by,
\bea
H_{reg} =
- {1\over2}\sum_{l,l'} |{\cal G}|^{1/2}
 (D_L)^a_l \big\{ |{\cal G}|^{-1/2}
 {\cal G}^{ab}_{ll'} (D_L)^b_{l'} \big\}          \ .
\ena
The F-P hamiltonian defines the lattice regularization of the
hamiltonian for 4D Euclidean quantum gravity in Ashtekar's
formalism\footnote{
In Refs.~\cite{RL}\cite{Loll}, similar lattice regularized hamiltonians
have been discussed without the contribution of $|{\cal G}|$.
}.
The second is the invariant measure which appears in the
path-integral representation of the regularized
expectation value of an observable (6)
after integrating out the canonical momentum variables, $(D_L)^a_l$.
\bea
{\cal D}U \equiv {\cal G}^{-1/2}\Pi_l dU_l   \ ,
\ena
where $dU$ denotes the left invariant Haar measure.
The measure is invariant under the general coordinate transformation in
superspace (19).

The third is the Schwinger-Dyson equation in this context.
It is given by~\cite{Na1}
\bea
< |{\cal G}|^{1/2} \Big\{ (D_L)^b_{l'}
\Big(  |{\cal G}|^{-1/2} {\cal G}^{ab}_{ll'} \Big)\Big\} T^a U_l
> = 0         \ ,
\ena
at an equilibrium limit, $\limt <{\dot U}> = 0 $.

In this note, I have pointed out that there exist a Langevin system for
3-geometries
which describes the time evolution in 4D Euclidean quantum gravity.
There have been some attempts to define 4D quantum gravity
in the context
of stochastic quantization~\cite{Sa}\cite{Ru}\cite{Naka}\cite{GS} by
introducing the stochastic time as the fifth time coordinate, however, the
philosophy of the present approach is different from these previous
works. The strategy I would like to adopt here is to characterize the
3-boundaries in the 4D spacetime manifold by using the solution of both
the momentum constraint and the Gauss law constraint to prepare
the initial distribution and
the observables. Then the remaining constraint is only the hamiltonian
constraint which may be realized at the infinite
stochastic time limit provided
there exists an equilibrium limit. This is actually the idea which we
have learned in stochastic quantization of matrix models in loop
space for 2D quantum gravity, where the stochastic time is
the proper Euclidean time and
the hamiltonian constraint is realized as the Virasoro constraint in
loop space~\cite{JR}\cite{Na1}. In order to work in this program, one has to
construct observables in the lattice regularization. It should be
noticed that the Wilson loop is not a spatially general coordinate
invariant observable in the present lattice regularization~\cite{Napre}.

In the Langevin equation (1) and (14),
the \lq\lq triad \rq\rq in Ashtekar's variables are
realized as noise variables, which presumably represents the
stochastic 3-geometries. The lattice regularized Langevin equation (14)
may provide a possible basis for numerical simulation in 4D quantum gravity.
The problem of the present scheme is
that the noise correlation in the basic Langevin equations, (1) and (14),
are not
positive definite. It would force the Langevin equations to be
complex. The field variables, though they are real in the Euclidean
Ashtekar's formalism, would also become complex in the time evolution.
The point would be main difficulty for the
numerical analysis in this scheme.
One way to deal with the problem may be to extend the Langevin equation
(1) to a class of more general gauge fixing. It is always possible by
multiplying the noise correlation in (1) with lapse function. Then one
may chose the lapse function so that the noise correlation keeps the
value to be positive definite. It is an open question if this gauge
fixing procedure, a choice of non-trivial lapse function,
make sense in the lattice regularization.
It is also an open question how we
introduce the cosmological term in (1) and (14). Apart from these questions,
the description with the Langevin equation has a topological feature
in the sense of Nicoli-Langevin map. Such a topological feature would
relax another difficulty, renormalizability of quantum gravity.
I hope that the approach is useful for deeper understanding of
quantum gravity.

\vs{10}
\noindent
{\it Acknowledgements}

The author would like to thank
J. Ambjorn, J.Greensite,
A. Krasnitz, H. B. Nielsen and
M. Weis for enlightening discussion and comments. He also wishes to
thank J. L. Petersen for continuous encouragement and all members
in high energy group at Niels Bohr Institute for hospitality.


\end{document}